\documentclass[prb,superscriptaddress,twocolumn]{revtex4}
\usepackage{graphicx,amssymb}
\begin{document}

\title{ Susceptibilities and spin gaps of weakly-coupled spin ladders }

\author{ S. Larochelle }
\affiliation{Department of Physics, Stanford University, Stanford, CA 94305}
\author{ M. Greven }
\affiliation{Department of Applied Physics, Stanford University, Stanford, CA 94305}
\affiliation{Stanford Synchrotron Radiation Laboratory,
    Stanford Linear Accelerator Center, Stanford, CA 94309}
\date{\today}

\begin{abstract}
We calculate the uniform and staggered susceptibilities of two-chain spin-$\frac{1}{2}$ 
Heisenberg ladders using Monte-Carlo simulations. We show that the gap
extracted from the uniform susceptibility and the saturation value of the staggered 
susceptibility are independent of the sign of the inter-chain coupling $J_\perp$ in 
the asymptotic limit $|J_\perp|/J \rightarrow 0$. Furthermore, we examine the 
existence of logarithmic corrections to the linear scaling of the gap with 
$|J_{\perp}|$.

\end{abstract}

\maketitle

Spin ladders are arrays of coupled spin chains and exhibit structures that 
interpolate between a single one-dimensional chain and the 
two-dimensional square lattice. 
Their properties are peculiar as the magnetic
spectrum is gapless only in the case of non-integer-spin ladders formed 
of an even number of chains. \cite{Dagotto96,White96} 
Spin-$\frac{1}{2}$ ladders are thus closely related to 
spin-S chains for which the excitations are gapped for integer spins and gapless
otherwise.\cite{Haldane83a}
An important question is the understanding of the transition between
the gapless chain and the gapped ladder (the Haldane phase). That issue
has been studied both analytically \cite{Troyer94,Shelton96} and numerically.
\cite{Hida95,Greven96}

In this paper, we report results of Monte-Carlo 
simulations on two-chain spin-$\frac{1}{2}$ ladders with antiferromagnetic coupling
along the chain direction and ferromagnetic or antiferromagnetic coupling 
across the rungs. We calculate the uniform and 
staggered susceptibilities and show that, in the asymptotically weak rung-coupling 
regime, these thermodynamic quantities are independent of the sign of the rung 
coupling. In this regime, our
results for the spin gap are consistent with the theoretically-predicted
existence of logarithmic corrections 
to the linear behavior $\Delta \sim |J_\perp|$.

The spin-$\frac{1}{2}$ Heisenberg ladder magnet is described by the following 
Hamiltonian:

\begin{equation}
  H=J \sum_{i=1}^N\sum_{n=1}^2{\bf S}_{i,n} {\bf \cdot S}_{i+1,n} + 
  J_{\perp}\sum_{i=1}^{N}{\bf S}_{i,1} {\bf \cdot S}_{i,2}
\end{equation}

\noindent where $i$ runs along the chains, $N$ is the length of the chains,
 $J$ is the antiferromagnetic  coupling along the chain direction and 
$J_{\perp}$ is the coupling between the two chains (the rungs of the ladder). We consider 
both antiferromagnetic ($J_{\perp}>0$) and ferromagnetic ($J_{\perp}<0$) rung couplings. 
The Hamiltonian Eq. (1) is investigated with the loop cluster algorithm 
with a discrete Euclidean time grid.\cite{Wiese94,Greven96}
Periodic boundary conditions are used along 
the chain direction as well as in the Euclidean time direction. The chain length is 
kept larger than six times the spin-spin correlation length \cite{csize}.
The Trotter number, which defines the discretization of the Euclidean
time axis, is kept larger or equal to $20/T$ where $T$ is the temperature.
We calculate the uniform and staggered susceptibilities, defined as: 
\begin{equation}
\chi_u(T)=\frac{1}{2 N T}\left<\left(\sum_{i=1}^{N}\sum_{n=1}^2 S_{i,n}^z\right)^2\right> 
\end{equation}
\begin{eqnarray}
&&\chi_s(T)= \frac{1}{2 N T}\nonumber\\  
&&\left<\left(\sum_{i=1}^{N}\sum_{n=1}^2 (-sgn(J))^i(-sgn(J_\perp))^nS_{i,n}^z\right)^2\right>
\end{eqnarray}

Studies at intermediate and large couplings show that finite-size effects, both in the 
Euclidean time direction and in the chain direction, are smaller than the statistical 
uncertainties for the two observables when the lattice dimensions were kept
larger than the previously defined limits. The same relations for the required
system size were assumed to apply in the small coupling regime.

\begin{figure}
  \includegraphics[width=8.8cm]{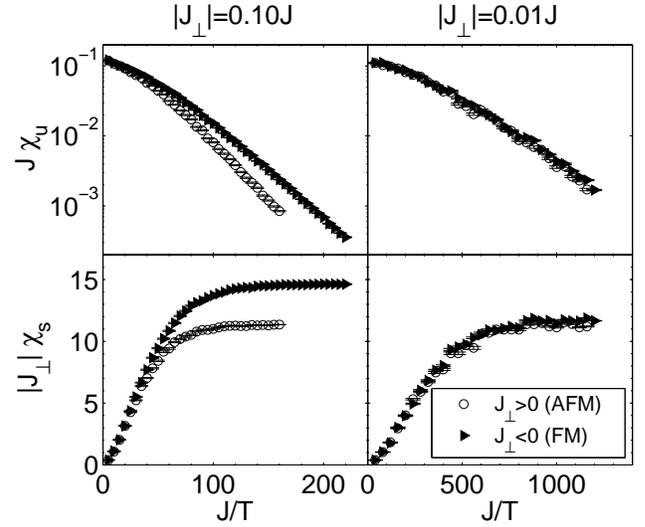}
  \caption{ Uniform (top) and staggered (bottom) susceptibilities
    as function of the inverse temperature for the spin-$\frac{1}{2}$ ladder. 
    Left (right) panels are for rung couplings $|J_\perp|=0.10J$
    ($|J_\perp|=0.01J$).
    \label{figsus}}
\end{figure}

Figure \ref{figsus} shows 
examples of the uniform susceptibility and staggered susceptibility for ferromagnetic and 
antiferromagnetic $J_{\perp}$. For $|J_\perp|=0.01J$ we find that 
the susceptibilities become independent of the sign
of $J_\perp$, consistent with theoretical expectations for two weakly-coupled
Heisenberg chains.\cite{Shelton96}

\begin{figure}
  \includegraphics[width=8.8cm]{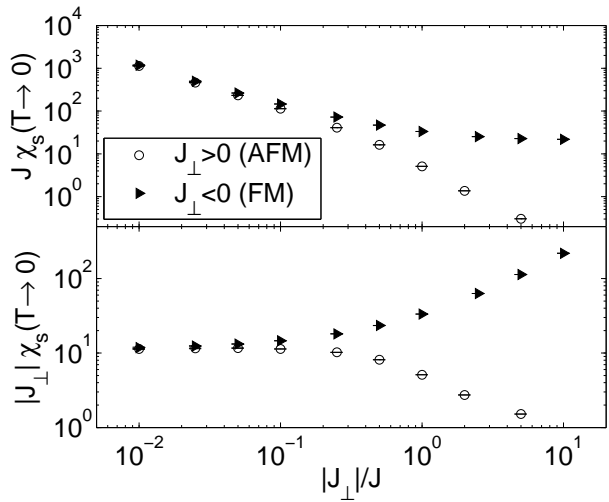}
  \caption{ Low-temperature staggered susceptibility as a function of the rung coupling. 
    \label{stag}}
\end{figure}

Figure \ref{stag} shows the dependence of 
the extrapolated zero-temperature staggered susceptibility 
on $J_{\perp}/J$. In the weak-rung-coupling regime, $\chi_s(0)$ varies linearly with 
$|J/J_{\perp}|$, as can be seen in the bottom panel of Fig. \ref{stag}. 
We find that $|J_{\perp}| \chi_s(0) =11.5(5)$ for $|J_{\perp}|/J \rightarrow 0$.  

\begin{figure}
  \includegraphics[width=8.8cm]{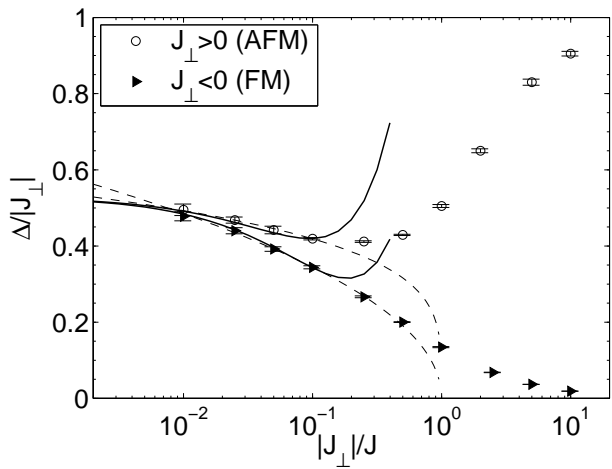}
  \caption{ Spin gap as function of the rung coupling for both ferromagnetic
    and antiferromagnetic rung couplings. The dashed lines are fits to 
    $D \left[ln(|J_\perp|/J)\right]^{E}$ with exponent $E$ of 0.5 and 0.23 for, 
    respectively, for ferromagnetic and antiferromagnetic rung coupling 
    (Eq. (\ref{log2}) and Eq. (\ref{log3}), respectively). The continuous lines 
    are fits to Eq. (\ref{log1}), with
    coefficients $A=0.53(2)$, $B=2.2(5)$, as well as $C=3.9(5)$ ($C=2.0(6)$)
    in the antiferromagnetic  
    (ferromagnetic) case.
    \label{gap}}
\end{figure}

The spin gap is extracted from the uniform susceptibility using the following low 
temperature form:\cite{Troyer94}

\begin{equation}
J \chi_u(J_\perp,T)\sim T^{-1/2}e^{-\Delta(J_{\perp})/T}
\end{equation}

\noindent where $\Delta$ is the gap. This limit is derived from a non-interacting 
magnon model with a quadratic band dispersion. The fits are performed for
$T\lesssim\Delta/3$, and the results are shown in Fig. \ref{gap}. The data for 
antiferromagnetic rung couplings ($J_\perp>0$) agree well with previously reported 
results.\cite{Greven96} In the weak coupling limit, the
gap deviates from a linear dependence on $J_{\perp}$. 
For the 
ferromagnetic rung coupling, values of the gap at intermediate rung couplings 
($0.1 \leq |J_\perp|/J<1$)  
are in
agreement with previous reports \cite{Watanabe94,Narushima95,comment}.

At very large ferromagnetic rung coupling, the spin-$\frac{1}{2}$ ladder is equivalent 
to a spin-1 chain with coupling $J/2$. The gap for the spin-1 chain was calculated to be 
$\Delta=0.41050(2)J$ \cite{White93} or $\Delta=0.4107(1)J$ \cite{Sorensen93}
by the density matrix renormalization group (DMRG) technique. In the present 
investigation, the spin-1 
chain gap was estimated at $0.406(3)J$, which agrees reasonably well with the DMRG value. 
At finite ferromagnetic coupling $J_{\perp}$, the ratio $\Delta/J_{\perp}$ increases
with decreasing value of $J_\perp$. At small rung coupling, the data for the 
antiferromagnetic and ferromagnetic rung coupling exhibit a similar trend.
Together with the behavior of the staggered susceptibility, this seems to confirm the
assertion \cite{Shelton96, Hosotani97} that the properties of a 
two-chain spin-1/2 spin ladder are
independent of the sign of $J_{\perp}$ in the weak coupling limit.

In the weak-coupling regime, the gap has a nearly linear dependence on the rung 
coupling $J_{\perp}$. Logarithmic corrections have been proposed to this linear 
dependence based on field theoretical considerations. Shelton, Nersesyan and Tsvelik
\cite{Shelton96} concluded that the gap follows the form:

\begin{equation}
\Delta=A|J_{\perp}|\left(1+B J_{\perp}ln\left(C\frac{|J_{\perp}|}{J}\right)\right) 
\label{log1}
\end{equation}

\noindent 
where  $A$, $B$ and $C$ are unknown constants. Totsuka and Suzuki \cite{Totsuka95}, 
on the other hand, suggested that the gap be described by

\begin{equation}
\Delta=D|J_{\perp}|\sqrt{ln\left(\frac{|J_{\perp}|}{J}\right)}
\label{log2}
\end{equation}

\noindent 
For $|J_{\perp}|/J \le 0.1$, Eq. (\ref{log1}) describes the gap well 
for both antiferromagnetic
and ferromagnetic rung couplings, while Eq. (\ref{log2}) satisfactorily describes only the 
ferromagnetic result. A modified form of Eq. (\ref{log2}) is necessary to obtain an acceptable 
description for antiferromagnetic rung couplings: 

\begin{eqnarray}
  \Delta=0.344(4)|J_{\perp}|\left(ln\left(\frac{|J_{\perp}|}{J}\right)\right)^{0.23(2)} 
    \label{log3}
\end{eqnarray}

\noindent 
Although Eq. (5) describes both results well below $|J_{\perp}| = 10^{-1}J$,
it would be necessary to know the gap at rung couplings 
well below $|J_{\perp}| = 10^{-2}J$ 
in order to conclusively establish 
that this form is asymptotically correct in the limit 
$|J_{\perp}|/J \rightarrow 0$.

In conclusion, we have calculated the uniform and staggered susceptibilities 
for the  
spin-$\frac{1}{2}$ two-chain Heisenberg ladder. We establish that the 
susceptibilities are 
independent of the sign of $J_\perp$ in the weak-coupling regime 
$|J_{\perp}|/J \ll 1$. In that regime, the staggered 
susceptibility is linear in $J/|J_{\perp}|$, while the excitation
gap is described by a logarithmic correction to the linear dependence on $|J_\perp|$.

We acknowledge valuable discussions with O.P. Vajk and Y. J. Kim. This work was 
supported by NSF CAREER Award No. DMR9985067
and by the DOE's Office of Basic Energy Sciences, Divisions of
Materials Sciences and Chemical Sciences, through the Stanford Synchrotron 
Radiation Laboratory.

\bibliography{ladder}

\begin{thebibliography}{16}
\bibitem[{csi()}]{csize}
\bibinfo{note}{While the correlation length was not measured in the present
  study, estimates were calculated from the scaling relations established in
  Ref. [6].}
\bibitem[{com()}]{comment}
\bibinfo{note}{We note that the coupling parameter $\lambda$ in
  Ref.\cite{Watanabe94,Narushima95} equals $-J_\perp/(2J)$ in our notation.}
\end{thebibliography}
\end{document}